\documentstyle[12pt, psfig]{article}
\def\reference{\parskip 0pt\par\noindent\hangindent 0.5 truecm}

\begin{document}
%
%
\title{Ubiquitous column density variability in Seyfert 2 Galaxies }
%


\author{G. Risaliti, $^{1,2}$
M. Elvis, $^{2}$ F. Nicastro $^{2}$
} 

\date{}
\maketitle

{\center
$^1$ Ossevatorio Astrofisico di Arcetri, Largo E. Fermi, 5, I-50125 Firenze, Italy\\risaliti@arcetri.astro.it\\[3mm]
$^2$ Harvard-Smithsonian Center for Astrophysics, 60 Garden Street, Cambridge, MA, 02138, USA\\[3mm]
}

%
\begin{abstract}
\noindent  We present a study of X-ray column density variability in Seyfert 2 galaxies.
We show that variations in N$_H$ are observed in almost all the objects with multiple
hard X-ray observations. Variation timescales (as short as a few months in several cases)
are not in agreement
with the standard scenario of a parsec-scale toroidal absorber.
We propose that the X-ray absorber in Seyfert galaxies is located much nearer to the
center than previously assumed, on the Broad Line Region Scale. An extension
of the model by M. Elvis (2000) can explain the observed variability.
We also show preliminary results of N$_H$ variability search inside single
X-ray observations, which suggest that variations can occur on timescales
of a few 10$^4$ seconds.
\end{abstract}

{\bf Keywords:}

\bigskip

%
%

\section{Introduction}
 Strong obscuration is observed in the hard (2-10 keV) X-ray spectra of Type 2 Active
Galactic Nuclei, where a photoelectric cut-off at energies E $>$ 1-2 keV indicates
the presence of a column density of absorbing gas N$_H > 10^{22}$ cm$^{-2}$.

The simplest geometry for this gas surrounding the nucleus is that of a torus
covering $\sim$ 80\% of the solid angle (in order to reproduce the 4:1 observed ratio
between unobscured and obscured AGN, Maiolino \& Rieke 1995). One of the unsolved
questions about this putative torus is its typical dimensions.
Detailed models have been proposed for both a 100 pc-scale torus (Granato et al. 1997)
and for a parsec-scale one (Pier \& Krolik 1992). Both models are supported by
observational evidence, so it is likely that both the components could be
present in AGN. Here we investigate the variability of the X-ray absorbing column density
in X-ray defined Seyfert 2s having column densities higher than $\sim 10^{22}$ cm$^{-2}$,
but less than 10$^{24}$ cm$^{-2}$ (in order to have a measurement of the photoelectric
cut-off in the 2-10 keV band). We collected all the data available in the literature
for Seyfert 2s and we complemented them with the analysis of unpublished data in the ASCA
and BeppoSAX public archives. In the following Sections we show the results and we show
that they can be explained within a consistent physical picture only assuming that
the absorber is located at a distance from the center typical of
the Broad Emission Line Region (BELR).


\section{Results}
We found that a sample of 25 sources were observed at least
twice in the hard X-rays. Out of these 25 sources, 22 show N$_H$ variability on
timescales from a few months to years (Fig. 1).
\begin{figure}
\centerline{\psfig{file=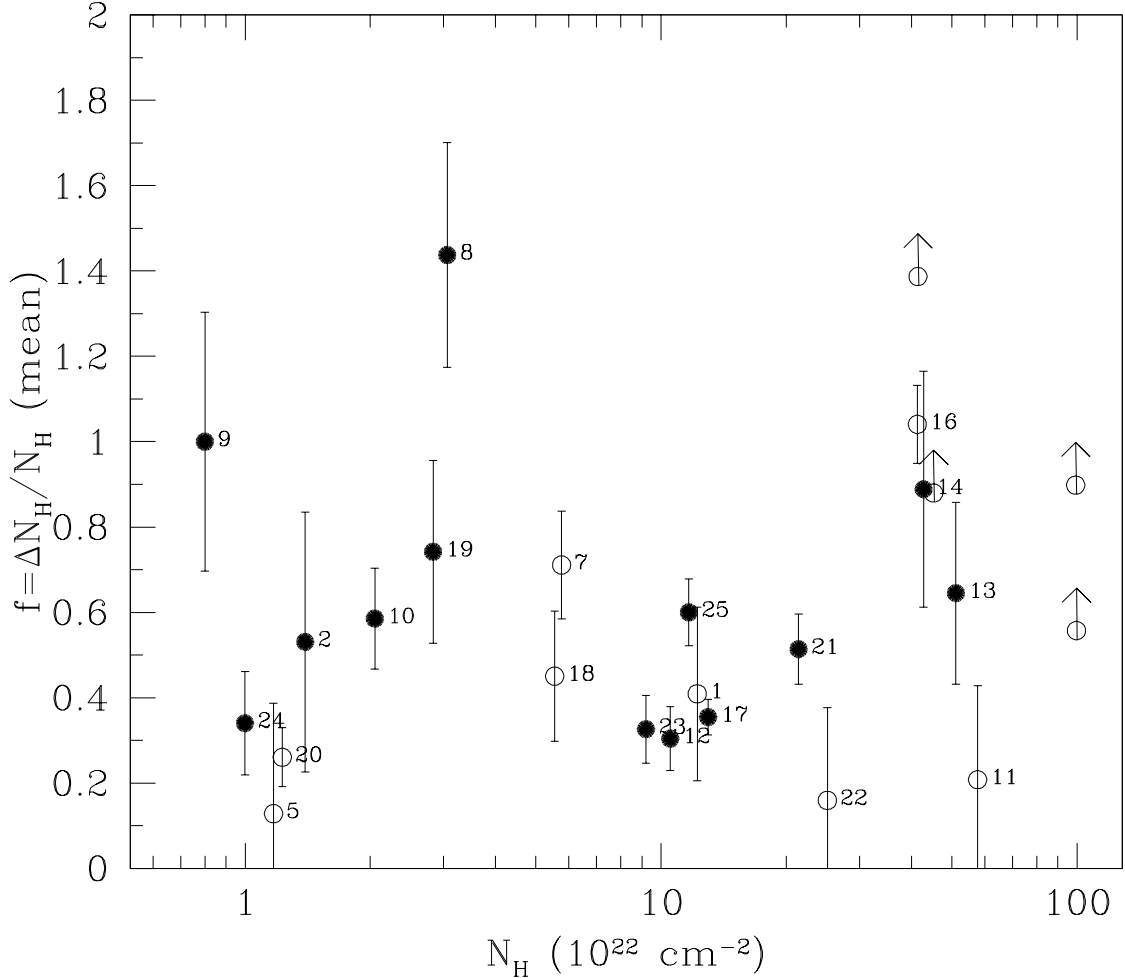,height=13cm}}
\caption{\footnotesize{Ratio between the column density variations, $\Delta N_H$ and the mean N$_H$ for all the Seyfert 2s
with multiple hard X-ray observations. Empty circles are for source observed
twice, filled circles are for sources with three or more observations.}}
\end{figure}
 The full detailed analysis of the 139 observations of these 25 sources
is described in Risaliti et al. 2001.

There are two physical reasons that can explain the variability of the absorbing
column density: variation in the ionization state of the absorber, due to variations
in the ionizing continuum, and variations of the amount of gas along the line of sight.
We ruled out the first possibility, since the N$_H$ variations are not correlated with
the flux variations (see Risaliti et al. 2001 for details). The second scenario -motions
in a clumpy medium- is the only one that can account for the observations.
\begin{figure}
\begin{center}
\psfig{file=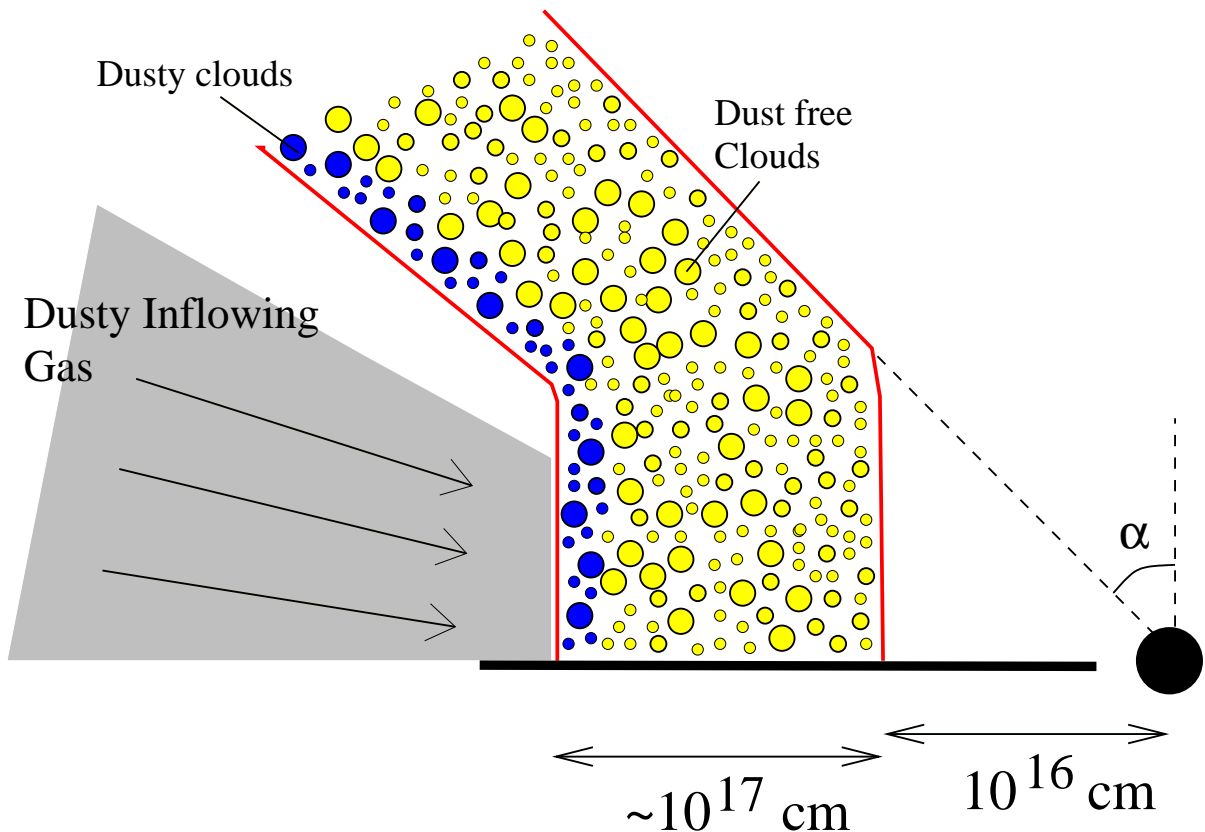,height=11cm}
\caption{\footnotesize{
A simple model, derived from Elvis (2000), which explains the X-ray absorption
properties of Seyfert 2s. The X-ray central emission is absorbed by the Broad Emission
Line
Clouds. The column density variability timescale is the average crossing time of a cloud
along
the line of sight. The covering factor of the funnel-shaped absorber (given by 1-cos$\alpha$)
should be of the order of 0.7-0.8, in order to reproduce the observed ratio
between type 2 and type 1 AGNs.}}
\end{center}
\end{figure}
However, the
results on the ubiquity of N$_H$ variability in Seyfert 2s, together with
the short (from 2 months to a few years) variability timescales, pose severe problems
for the standard torus model. We can idealize the situation by assuming the typical
timescale of
variation, $t$, to be the crossing time of a discrete cloud across the line of sight.
Assuming
that the absorption is due to spherical clouds moving with Keplerian velocities, the
distance
from the central black hole of mass $M_\bullet$ is given by:
\begin{equation}
R \sim 3 \times 10^{16} \frac{M_\bullet}{10^9M_\odot}~(\frac{\rho}{10^6 {\rm
cm}^{-3}})^2~(\frac{t}
{{\rm 5~ Msec}})^2~(\frac{N_H}{10^{22}{\rm cm}^{-2}})^{-2} ~{\rm cm}
\end{equation}
where $\rho$ is the density of the cloud. The black hole mass and the cloud density have
been
normalized to extreme values for a putative torus in order to obtain the greatest
distance.

The distance we obtained is typical for the BELR. Even if many parameters are poorly
constrained, it is not possible to obtain a value of $R > 10^{18}$ cm
within a
consistent physical scenario. Therefore, the parsec-scale torus model is not able to
explain our
data.
 An alternative scenario, within the standard AGN model (Antonucci 1993), is that the X-ray
absorber is located in the BELR, much nearer to the central black hole than the standard
torus.
If we assume that the broad line clouds are responsible for the absorption in the X-rays,
we can
find a consistent combination of the parameters in the previous equations, with higher
cloud densities ($\rho \approx 10^9 cm^{-3}$) and shorter variability timescales ($t
\approx 3$
days). We note that such timescales are not ruled out by our data, since we cannot investigate
variations shorter than
the time interval between two observations of the same
source.

An absorber which is very compact (as required by our data) and axisymmetric (as required
by
the arguments supporting the unified schemes) can be easily obtained by extending the wind
model by
Elvis (2000, and these proceedings). In this model most of the phenomenology of type 1 AGNs
is
explained through a two-phase wind arising from the accretion disc. The cold phase of the
wind
is formed by the Broad Emission Line Clouds (BELC). A simple extension of this model could
be that in type
2 AGNs the wind is thicker, and the BELC cover all lines of sight through the wind, as
illustrated in Fig. 2. The external part of the wind can well be cold enough for dust to
survive, therefore this absorber can also explain the optical properties of Seyfert 2s.
Interestingly, the average
dust-to-gas ratio predicted by this model, assuming that the external part of
the wind has a standard ISM composition, is lower than Galactic, in agreement with recent
findings (Maiolino et al. 2001, Risaliti et al. 2001)


\section{Conclusions, and future work}

Variability of X-ray absorbing column density appears to be an ubiquitous property in Seyfert 2
galaxies. Variation timescales can be as short as a few months. We have shown that these data
rule out an X-ray absorber on a parsec scale. Instead, we propose that
the Broad Emission Line Clouds, in an axisymmetric distribution, like in the model of Elvis
(2000) are responsible for X-ray absorption. If this is the case, variations in N$_H$ are
expected on timescales of days. Our variability study is limited by the shortest time between
two observations of the same source. However, our work can be significantly improved looking for
N$_H$ variations inside single, long observations of the brightest sources in our sample.
This work is still in progress. However we can show some preliminary,promising results.
In Fig. 3 we plot 3 light curves for three different energy intervals of a BeppoSAX observation
of the Seyfert 2 ESO 103-G35. 
\begin{figure}[h]
\centerline{\psfig{file=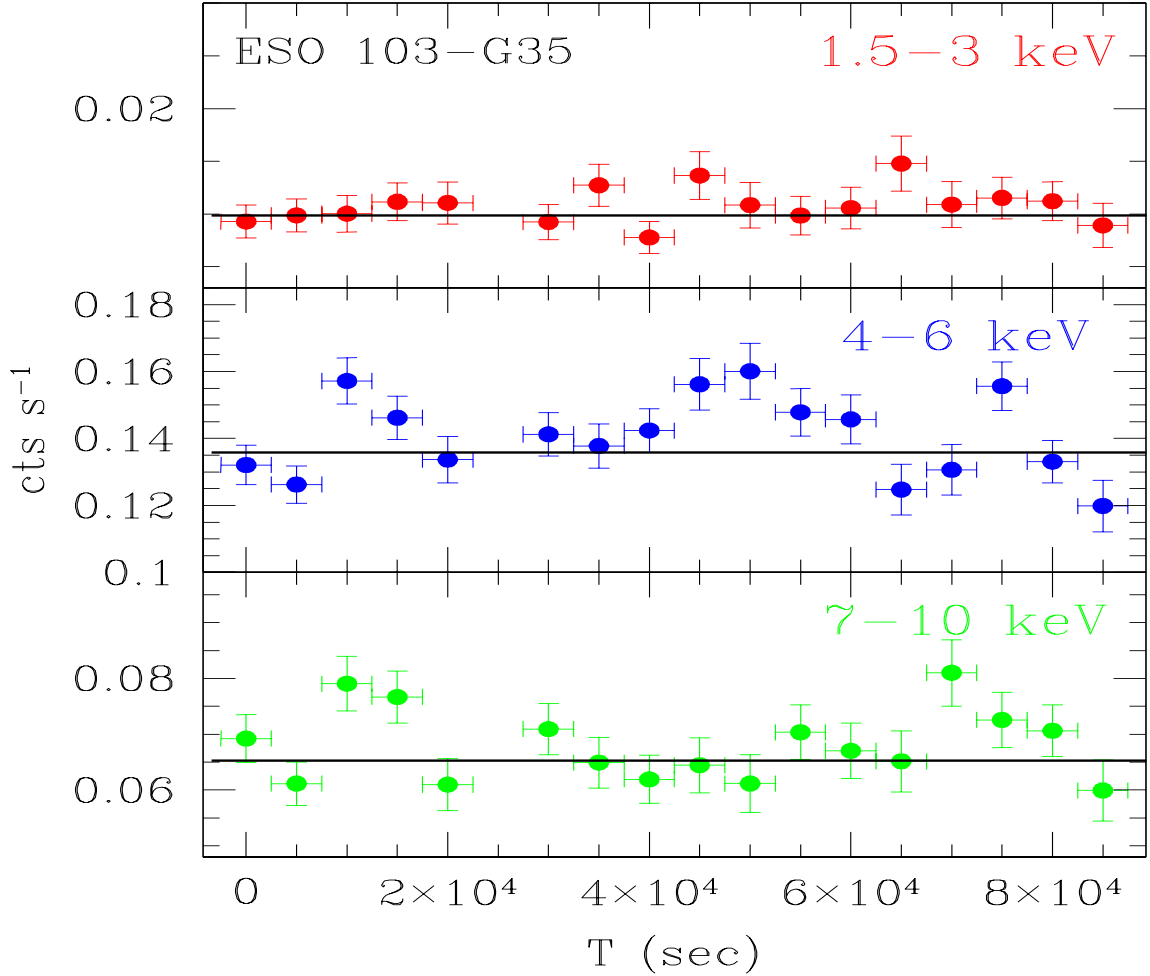,height=13cm}}
\caption{\footnotesize{X-ray light curves of a BeppoSAX observation of ESO 103-G35. The
excess between 40 and 60 ksec in the second is suggestive of an N$_H$
variation. A subsequent spectral analysis has shown that this variation is
significant at a 5 sigma level of confidence.}}
\end{figure}

\noindent The intervals are chosen is order to have the photoelectric
cut-off at the separation energy between the first two lightcurves. Therefore,
in case of variations of N$_H$,
we expect that one of the first two light curves varies, while the
other two remain constant (the third energy interval is little affected by small N$_H$
variations). Instead, if we have a flux variation, we expect both the second and third light
curve vary, while the first remains constant, since the emission at energies lower that the
 cutoff is mainly due to an extended/reprocessed component.
It is clear from Fig. 3 that in the interval between 40 and 60 ks the light curves vary in
a way suggesting N$_H$ variability. To check if this is indeed the case, 
we extracted two spectra, one in the 40-60 ks interval,
and the other in the remaining time intervals. We performed a careful fit of these spectra,
and concluded that the N$_H$ measurements differ by $\sim 3\times 10^{22}$ cm$^{-2}$, at a
5$\sigma$ level of confidence. This N$_H$ variation on a timescale of $\sim$ 20 ks strongly
suggests that the X-ray absorber is very close to the central black hole, in agreement with the
model we have proposed.


\medskip
\noindent This work was supported in part by NASA grant NAG5-4808.

 \section*{References}

\reference Antonucci, R. R. 1993, ARA\&A, 31, 473
\reference Elvis, M. 2000, ApJ, 545, 63
\reference Granato, G., Danese, L., \& Franceschini, A. 1997, ApJ, 486, 147
\reference Maiolino, R., \& Rieke, G.H. 1995, ApJ, 454, 95
\reference Maiolino, R., Marconi, A., Salvati, M., Risaliti, G.,
Severgnini, P., Oliva, E., La Franca, F., \& Vanzi, L. 2001, A\&A, 365, 28
\reference Pier, E.A., \& Krolik, J.H. 1992, ApJ, 401, 99
\reference Risaliti, G., Marconi, A., Maiolino, R., Salvati, M.,
\& Severgnini, P. 2001, A\&A, 371, 37

\end{document}